\documentstyle[11pt]{article}
\newcommand{\llg}{Lorentz lattice gas }
\newcommand{\NC}{$\bf NC$}
\newcommand{\Poly}{$\bf P$}

\begin{document}

\title{The Computational Complexity of the Lorentz Lattice Gas}

\author{J. Machta and K. Moriarty\\
Department of Physics and Astronomy\\
University of Massachusetts\\
Amherst, Massachusetts 01003-3720\\
}
\maketitle

\begin{abstract}
The \llg is studied from the perspective of computational
complexity theory. It is shown that
using massive parallelism, particle trajectories can be
simulated in a time that scales logarithmically in
the length of the trajectory.  This result characterizes the ``logical
depth" of the \llg and allows us to compare it to other models in
statistical physics. 

\end{abstract}

\section{Introduction}

The \llg is a simple but non-trivial model of single particle
transport in a quenched random environment. Ernst and
collaborators~\cite{ErVe89a,ErVe89b,vBEr,ErDoJa} have used the \llg
to examine a number of interesting questions in nonequilibrium
statistical physics.  In this paper we will examine this model from
the perspective of its computational complexity.  Though this study
may aid in the design of simulations, our primary motivation
is to provide a new characterization of this model. 

The \llg can be described in the following way.  A single
particle moves with unit speed on a lattice.  Some of the
sites of the lattice are scatterers.  In the absence of
scatterers the particle moves in a straight line along one
of the lattice directions but when it encounters a scatterer
its velocity may be changed.  Two broad classes of models can be
distinguished, depending on whether the motion of the particle is
deterministic or stochastic.  For the stochastic \llg each
collision is a random event whereas for
the deterministic model the effect of the scatterer depends only on
the incoming velocity.  The behavior of these two classes is quite
different~\cite{CoWa} but the computational approach described
below treats both on an equal footing.

Computational complexity is the branch of theoretical
computer science that seeks to quantify the resources
required to solve problems.  In this study, we focus on
parallel computation and ask the following question. 
Suppose we are supplied with a massively parallel computer
and we are asked to generate representative \llg
trajectories of length $t^*$.  How
much time will this take?  

First consider the time requirements for a straightforward and
physically motivated parallel approach.  We could connect the
processors as a cellular automaton with every site of the lattice
controlled by a processor.  When a particle arrives at a given site the
local processor passes the particle to the appropriate neighboring
site.  It is clear that this approach will require a computational
time that scales as the physical time $t^*$ and uses $O(L^d)$
locally connected processors where $L$ is the system size and $d$ the
dimension.  Suppose we are willing to employ many more processors 
and have them non-locally connected.  Could we obtain a computation
speed that is qualitatively faster? To answer this question and place
it in context we must first briefly discuss computational complexity
theory.

\section{Computational Complexity}
\label{se-complexity}

Due to space limitation it is impossible to provide more than the
most superficial introduction here.  The reader can find further
information and details in a number of texts~\cite{hu79,lp81} and
monographs~\cite{GrHoRu,Jo90,GaJo,GiRy}.  A somewhat more detailed 
discussion of applications to problems in statistical physical can
be found in Refs.\ \cite{MaGr,MaGr96}.

Complexity theory typically considers decision problems (i.e.\
problems with YES or NO answers) of varying size $N$.  The size of
the problem is defined as the number of bits required for the problem
input.  Complexity theory is usually expressed in terms of space
and time requirements on Turing machines, however,  equivalent
formulations in terms of parallel computation (see for
example~\cite{GiRy,Ve,BaIm}) are better suited to the (extensive)
problems of statistical physics.   The standard theoretical model of
parallel computation is the `P-RAM', consisting of a number of
processors all running the same program and all connected to a global
random access memory that can be accessed in unit time.  The measure
of computation time in the P-RAM model is unphysical because of the
assumption of unit access time for any processor to any memory
element.  Nonetheless, this measure of time is
interesting because it reveals the ``logical depth'' of
the computation.  The notion of logical depth is defined within an
equivalent parallel computational model: ``families of Boolean
circuits.''  A Boolean circuit is a collection of AND, OR and NOT
gates connected together without feedback.  A circuit has $N$ inputs
and a single output.  The {\em depth}\/ of the circuit is the longest
path from an input to the output while the {\em size}\/ is the number of
connecting wires. To solve a problem with inputs of varying sizes, a
family of circuits, one for each $N$, is required.  For
parallel computation, the important resources are hardware (processors
and memory for the P-RAM or, equivalently, the size of a Boolean circuit
that solves the problem) and parallel time or, equivalently the depth
of the corresponding Boolean circuit.  The focus of this paper is on the
resource of parallel time (logical depth).

A concrete example may help clarify the notion of P-RAM computation
and how it can be applied in statistical physics. 
Consider an ordinary random walk on a lattice.  At each time step the
walker chooses one of the $2d$ lattice directions.  Suppose the
objective is to generate a sample random walk of length $T$ on a
lattice of size $L^d$ with $L=O(T)$.  Let ${\bf r}(t)$ be the position
of the walker at time $t$ and let $\delta{\bf r}(t)$ be the random
displacement of the walker at time $t$.  Then ${\bf r}(t)={\bf
r}_0+\delta{\bf r}(1)+\cdots+\delta{\bf r}(t)$ where ${\bf r}_0$ is the
initial position of the walker. The displacements are chosen from
the unit vectors of the lattice.  Random walk simulations using this
approach can be recast as the following decision problem by taking the
random displacements as inputs: \\

\noindent 
{\bf Random Walk} (dimension $d$)\\ 
{\bf Given:} System size $L$ and duration $T$, displacements
$\{\delta{\bf r}(t) | t=1,\ldots,T\}$ and an initial position ${\bf
r}_0$. A designated position ${\bf r}^*$
and time $t^*$.\\
{\bf Problem:}  Is the particle at position 
${\bf r}^*$ at time $t^*$?\\

\noindent Note that the problem size $N$ here is $O(T)$ because of the
list of random displacements.  Of course, the decision problem is only
part of the full problem of simulating a random walk.  The first
task is to generate the random displacements by some means. Leaving
aside the many interesting questions related
to generating pseudo-random numbers, we assume that
the required random numbers can be obtained in parallel in unit time.
Furthermore, we are likely to be interested in the whole trajectory,
not just whether the walk visits a certain site at a certain time.  To
obtain the full trajectory up to time $T$ we solve in parallel $T L^d$
decision problems.  This approach is extravagantly wasteful of hardware
but does not increase the parallel time.  Thus the overall logical
depth of generating a random walk by the above method is essentially the
same as for the above random walk decision problem.  

The random walk decision problem is
reduced to summing ${\bf r}_0$ and the $t^*$ displacements
(additional complications arise if the walk encounters the boundary
but they do not affect the conclusion).  The result is then compared
with ${\bf r}^*$.  Since the problem size scales as $T$, the problem
may be solved on a conventional sequential computer in a time that is
{\em polynomial} (in this case nearly linear) in the problem size.  A
problem that can be solved on a sequential computer (or Turing
machine) in a time that is bounded by a polynomial in the problem size
is said to be in the class \Poly.  The random walk problem is thus
placed in the complexity class \Poly.

Given a P-RAM however we can add $n$ numbers much more quickly.  The
numbers are  assumed to be in the first $n$ elements of the global
memory.  The procedure requires $n/2$ processors.  In the first step,
each processor takes a distinct pair of numbers from memory, adds them
and returns the sum to memory leaving $n/2$ partial summands.  In
subsequent steps processors are again assigned to add pairs of numbers
returning them to memory.  Since the number of partial summands is
halved in each step, only $O(\log n)$ steps are required to complete
the summation.  Here the problem size $N$ is proportional to $n$.

A problem that can be solved on a P-RAM using polynomially many 
(i.e.\ $N^{O(1)}$)
processors in polylog (i.e.\ $\log^{O(1)} N$) time is said to be in the
class \NC.  Note that ${\bf NC} \subseteq {\bf P}$ since polynomially
many processors running for polylogarithmic time can be simulated by a
single processor running for polynomial time. Problems in \NC\/
do not have much logical depth.  

The above
arguments show the random walk problem is in the class \NC. Although
the random walk takes physical time $T$ to unfold, it can be generated
using only $O(\log T)$ logical steps.  This result sets only an upper
bound on the complexity of sampling random walks since we have not
excluded the possibility that a different sampling method may be faster
than adding up $T$ numbers.

\section{Fast parallel algorithm for the \llg}

The particle trajectory in the \llg can be viewed as a path on a
$d+d+1$ dimensional lattice.  A point on this lattice takes
the form $({\bf r},{\bf c}, t)$ where ${\bf r}$ is a
$d$-dimensional position vector with each co-ordinate taking
integer values from 0 to $L$\/ and ${\bf c}$ is the
$d$-dimensional velocity vector.  The particle is constrained
to move along lattice directions with unit speed so the
velocity vector has exactly one nonzero component that may be
either $+1$ or $-1$.  The time $t$ can take any
integer value from $0$ to $T$ and suppose that $L=O(T)$.  

The particle travels in a straight line along lattice
directions unless it encounters a fixed scatterer.  The set
$S=\{{\bf s}_1,\ldots,{\bf s}_M\}$ specifies the scatterer
locations with each ${\bf s}$ being a distinct lattice site.  When a
particle encounters a scatterer its velocity is modified.  Associated
with each scatter and each time is an impulse variable $\delta_k(t)$
that determines how the velocity is changed if the
particle is at the $k^{\rm th}$ scatterer at time
$t$.  $\delta$ may take values $0,1,\ldots,2d-1$
where `0' indicates forward scattering, `1' backscattering and the
other values, scattering into one of the $2(d-1)$
transverse directions (defined relative to the incoming
velocity).  We will also use the operator notation 
$R[\delta_k(t)]{\bf c}$ for the new velocity after the scattering
event.  Let $\Delta$ be the set of $MT$ impulses.  The \llg decision
problem is stated as follows:\\

\noindent 
{\bf Lorentz Lattice Gas} (dimension $d$)\\ 
{\bf Given:} System size $L$, 
duration $T$, scatterer locations $S$,
impulses $\Delta$ and an initial phase-space point $({\bf
r}_0,{\bf c}_0 )$. A designated phase-space point $({\bf
r}^*,{\bf c}^*)$ and time $t^*$.\\  
{\bf Problem:}  Is the particle at phase-space point 
$({\bf r}^*,{\bf c}^*)$ at time $t^*$?\\

The \llg problem is clearly in the class \Poly\/ but it is not
obvious that it is in \NC. The sequence of
scattering events appears to be history dependent and the simple
approach used for the ordinary random walk will not succeed here. 
Nevertheless, we now show that the \llg decision problem is in the
class \NC.  We do this by sketching a P-RAM algorithm that runs in time
$\log N$ with $N$ the problem size.  The first step in the
algorithm is the construction of a directed ``phase-space-time'' graph,
${\cal G}=({\cal V},{\cal E})$ from the scatterer locations and
impulses. ${\cal G}$ is independent of the initial and final
phase-space points.  The set of vertices ${\cal V}$ of ${\cal G}$ are of
the form $({\bf r},{\bf c}, t)$, bounded in the obvious way by $L$ and
$T$.  The directed edges ${\cal E}$ of ${\cal G}$ correspond to the
possible dynamics of the particle.  We write a  directed edge from
vertex $v_1$ to vertex $v_2$ as $\langle v_1,v_2 \rangle$. The rules for
constructing ${\cal E}$ are given below:   \begin{enumerate}  
\item[Free motion:] If ${\bf r}$ is not a scatterer position then for
each $0 \leq t < T$ and each {\bf c},\\ $\langle ({\bf r},{\bf c}, t),({\bf r}+{\bf c}
,{\bf c}, t+1) \rangle \in {\cal E}$.   
\item[Scattering:] If, for some
$k$, ${\bf r}=s_k$ then for each $0 \leq t < T$ and each {\bf c},\\ 
$\langle ({\bf r},{\bf c}, t),({\bf r}+
R[ \delta_k(t)]{\bf c},R[ \delta_k(t)]{\bf c},
t+1) \rangle \in {\cal E}$.
\item[Boundary:]The boundary conditions must be built into
${\cal G}$.  This is done in the obvious way.  
\end{enumerate}

The number of
vertices in ${\cal G}$ is $|{\cal V}|=2dTL^d$. The directed edges can
be described by a (sparse) $|{\cal V}|\times|{\cal V}|$ matrix with
ones representing edges and zeroes otherwise.  This matrix can be
entered into memory in a few parallel steps using $|{\cal V}|^2$
processors, one for each possible edge.

It is clear that a directed path exists from $({\bf
r}_0,{\bf c}_0,0 )$ to $({\bf
r}^*,{\bf c}^*,t^*)$ if and only if a particle with initial
position and velocity $({\bf r}_0,{\bf c}_0 )$ will
arrive at ${\bf r}^*$ with velocity ${\bf c}^*$ at time
$t^*$.  Thus, the computational task has been reduced to
determining whether such a path exists.  It turns out
that this {\em graph accessibility problem} (GAP) is a
well studied problem in complexity theory.  It is known
that GAP is in \NC~\cite{Jo90}.   

Here is a straightforward \NC\/ algorithm for GAP.  This approach is
not necessarily optimal but has a simple ``renormalization group''
flavor.  Consider a directed graph with $n$ vertices,
$\{v_1,\ldots,v_n\}$.  The adjacency matrix is an $n \times n$ matrix
of 0's and 1's. Each entry represents a possible directed edge,
$(v_i,v_j)$ where a `1' means the edge exists. The output of the
algorithm is the connectivity matrix $C(v,v^\prime)$, also an $n \times
n$ matrix of 0's and 1's where now a `1' means that there is a directed
path from $v$ to $v^\prime$. Initially the connectivity matrix is equal
to the adjacency matrix. The algorithm requires $n^3$ processors, one
for each triple $(v_i,v_k,v_j)$ of vertices.  The processor assigned to
$(v_i,v_k,v_j)$ reads the current connectivity matrix elements
$(v_i,v_k)$, $(v_i,v_j)$ and $(v_k,v_j)$.  The processor then answers
the question of whether, on the basis of its information, there is a
directed path from $v_i$ to $v_j$. Explicitly, if either $C(v_i,v_j)=1$
or both $C(v_i,v_k)=1$ and $C(v_k,v_j)=1$ then processor
$(v_i,v_k,v_j)$  writes a 1 to matrix element
$C(v_i,v_j)$ otherwise it does nothing.   Initially the connectivity
matrix knows of all pairs of vertices connected by paths of length one
(i.e.\ the edges).  After one step of the algorithm, the connectivity
matrix knows of all pairs of vertices connected by paths of length two
or less.  After the $k^{\rm th}$ step, the connectivity matrix knows of
all directed paths of length $2^k$ or less.  The algorithm stops as
soon as no changes are made to the connectivity matrix. If two points
are connected, there must be a path of length less than $n$, thus the
algorithm is assured of halting after $O(\log n)$ parallel steps. 
Since the time is logarithmic and the number of processors is polynomial
in the problem size it follows that GAP is in the complexity class
\NC.  The GAP subroutine is the most complex part of solving the \llg
problem so it follows that the \llg is also in \NC.  The parallel
approach used here is quite similar to that employed for several
growth models (e.g.\ the solid-on-solid model)~\cite{MaGr}.

\section{Discussion}

We have seen that both the ordinary random walk and the \llg yield
natural decision problems that can be solved in logarithmic parallel
time though the approach needed for the \llg is considerably more
sophisticated.  There is a fine-grained distinction between the
complexity of the two problems.  The GAP algorithm used in the
solution of the \llg  implicitly assumes the ``concurrent read
concurrent write'' (CRCW) P-RAM model.  The CRCW P-RAM allows all
processors to attempt to write to a single memory element at the same
time.  The CRCW P-RAM model is equivalent to families of Boolean
circuits with gates having arbitrary {\em fan-in}.  The fan-in of a
gate is the number of incoming logical values.   If we restrict our
gates to fan-in two, then an extra power of $\log N$ appears in the
parallel time requirement for GAP.  On the other hand, adding $N$
numbers does not use this feature of arbitrary fan-in.  The conclusion
of all this is that the \llg algorithm is slightly more complex than
the random walk algorithm.  Specifically, the \llg problem is in the
class ${\bf AC}^1$ (solvable by a family of arbitrary fan-in circuits
of depth $O(\log N)$ and polynomial size) while the random walk problem
is in the class \NC$^1$ (solvable by a family of fixed fan-in circuits
of depth $O(\log N)$)~\cite{Jo90}.  These and the other complexity
classes described above are easily seen to be related as
follows~\cite{Jo90}:    \begin{equation}
 {\bf NC}^1 \subseteq {\bf AC}^1 \subseteq {\bf NC} \subseteq {\bf P}
\end{equation}
Although it is widely believed that each of these inclusions is
strict, these conjectures have not been proved.  

Even assuming the above conjectures, our results provide
only upper bounds on the complexity of sampling the trajectories of
the random walk and \llg models.  Although it seems unlikely, there may
be a wholly different sampling method associated with a less complex
decision problem.  Furthermore, if one wishes to sample 
the locations of the particle at a given time and not whole
trajectories, less complex approaches may be available if the spatial
probability distribution is known.  The simplest example of this is an
ordinary random walk in a bounded space where, after a long time, all
sites are visited with equal probability.  Sampling the locations of the
walker in this limit is trivial and requires only constant parallel
time.

The logical depth of the \llg is roughly comparable to that of a
variety of growth models in statistical physics such as the Eden model
and invasion percolation~\cite{MaGr}, all of which are associated with
problems in the complexity class \NC.  These models
display less logical depth than diffusion limited aggregation which
apparently requires more than polylogarithmic parallel
time to simulate~\cite{MaGr96}.  It would be interesting to extend
this study to related dynamical systems such as the continuum
Lorentz gas and many-body lattice gases.

\section*{Acknowledgements}

This work was supported in part by National Science Foundation
Grant~DMR-9311580.  We thank Ray Greenlaw for useful comments.

\end{document}